\documentclass[12pt]{article}

\usepackage{amssymb}
\makeatletter \@addtoreset{equation}{section} \makeatother

\def\ftoday{{\sl {Le \number\day \space\ifcase\month
\or janvier\or f\'evrier\or mars\or avril\or mai \or juin\or juillet\or
ao\^ut\or septembre\or octobre \or novembre \or d\'ecembre\fi\space
\number\year}}}
\def\ptoday{{\sl {\number\day \space de\space \ifcase\month
\or janeiro\or fevereiro\or mar{\c c}o\or abril\or maio \or junho\or
julho\or agosto\or setembro\or outubro \or novembro \or dezembro\fi\space
de\space \number\year}}}
\def\gtoday{{\sl {Den \number\day. \ifcase\month
\or Januar\or Februar\or M\"arz\or April\or Mai \or Juni\or Juli\or
August\or September\or Oktober \or November \or Dezember\fi\space
\number\year}}}
\def\today{{\sl {\ifcase\month
\or January\or February\or March\or April\or May \or June\or July\or
August\or September\or October \or November \or December\fi
\space\number\day,\space
                                            \number\year}}}

\newcommand{\XI}{\XI}

\newcommand{\es}{\\[3mm]}

\newcommand{\sla}{\raise.15ex\hbox{$/$}\kern -.57em}
\newcommand{\Sla}{\raise.15ex\hbox{$/$}\kern -.70em}

\newcommand{\complex}{{\kern .1em {\raise .47ex
\hbox {$\scriptscriptstyle |$}}
    \kern -.4em {\rm C}}}
\newcommand{\real}{{{\rm I} \kern -.19em {\rm R}}}
\newcommand{\rational}{{\kern .1em {\raise .47ex
\hbox{$\scripscriptstyle |$}}
    \kern -.35em {\rm Q}}}
\renewcommand{\natural}{{\vrule height 1.6ex width
.05em depth 0ex \kern -.35em {\rm N}}}

\newcommand{\twiddle}{\lower.9ex\rlap{$\kern -.1em\scriptstyle\sim$}}

\newcommand{\eq}{\begin{equation}}
\newcommand{\eqn}[1]{\label{#1}\end{equation}}
\newcommand{\eea}{\end{eqnarray}}
\newcommand{\eqa}{\begin{eqnarray}}
\newcommand{\eqan}[1]{\label{#1}\end{eqnarray}}
\newcommand{\ba}{\begin{array}}
\newcommand{\ea}{\end{array}}
\newcommand{\eqac}{\begin{equation}\begin{array}{rcl}}
\newcommand{\eqacn}[1]{\end{array}\label{#1}\end{equation}}

\begin{document}

{\hfill\parbox{45mm}{{hep-th/xxxxxxx\\
CBPF-NF-xxx/yy }} \vspace{3mm}

\begin{center}
{\Large \bf The $N=2$ and $N=4$ Supersymmetric Extensions of the Lorentz- and
\es CPT-Violating Term in Abelian Gauge Theories}
\end{center}
\vspace{3mm}

\begin{center}{\large
Wander G. Ney$^{a,b,c,*}$, J. A. Helayel-Neto$^{b,c,**}$ and
Wesley Spalenza$^{b,c,d,***,}$\footnote{Supported by the Conselho
Nacional de Desenvolvimento Cient\'{\i}fico e Tecnol\'{o}gico CNPq
- Brazil.} } \vspace{1mm}

\noindent 
 $^{a}$Centro Federal de Educa\c{c}\~{a}o Tecnol\'{o}gica de Campos (CEFET) RJ-Brazil\\
$^{b}$Centro Brasileiro de Pesquisas F\'{i}sicas (CBPF) RJ-Brazil\\
$^{c}$Grupo de F\'{i}sica Te\'{o}rica Jos\'{e} Leite Lopes (GFT)\\
$^{d}$Scuola Internazionale Superiore di Studi Avanzati (SISSA)
Trieste-Italy.
\end{center}
\vspace{1mm}

{\tt E-mails: *wander@cbpf.br, **helayel@cbpf.br,
***spalenza@sissa.it}


\begin{abstract}
In this work, we propose the $N=2$ and $N=4$ supersymmetric
extensions of the Lorentz-breaking Abelian Chern-Simons term.
We formulate the question of the Lorentz violation in $6$ and $10$
dimensions to obtain  the bosonic sectors of $N=2-$
and $N=4-$ supersymmetries, respectively. From
this, we carry out an analysis in $N=1-\,D=4$ superspace and, in terms of $%
N=1-$ superfields, we are able to write down the $N=2$ and $N=4$
supersymmetric versions of the Lorentz-violating action term.
\end{abstract}

\section{Introduction}

The formulation of physical models for the fundamental interactions in
the framework of quantum field theories for point-like objects
is based on a number of principles, among which Lorentz covariance and
invariance under suitable gauge symmetries. However, mechanisms for the breakdown of these
symmetries have been proposed and discussed in view of a number of
phenomenological and experimental evidences \cite{1,1b,1c,1d,1e}.
Astrophysical observations indicate that Lorentz symmetry may be
slightly violated in order to account for anisotropies. Then, one
may consider a gauge theory where Lorentz symmetry breaking may be
realized by means of a term in the action. A Chern-Simons-type
term may be considered that exhibits a constant background
four-vector which maintains the gauge invariance but breaks down
the Lorentz space-time symmetry \cite{1}.

In the context of supersymmetry (SUSY), the issue of Lorentz violation
has been considered in the literature in different formulations: in
ref. \cite{3a}, supersymmetry is presented by introducing a suitable
modification in its algebra; in ref. \cite{3,3c}, one achieves the
$N=1-$SUSY version of the Chern-Simons term by means of the conventional
superspace-superfield formalism; in ref. \cite{3b}, the authors
adopt the idea of Lorentz breaking operators. More particularly,
considering the importance of extended supersymmetries in
connection with gauge theories, we propose in this work an $N=2$
and an $N=4$ extended supersymmetric generalization of the
Lorentz-breaking Chern-Simons term in a
4-dimensional Minkowski background. We start off with the Chern-Simons term in $%
(1+5)$ and $\left( 1+9\right) $ space-time dimensions and adopt
a particular
dimensional reduction method, see \cite{2}, to obtain the bosonic sector in $%
D=(1+3)$ of the $N=2$ and $N=4$ supersymmetric models, respectively. This
is possible because in $N=1,D=6$- and $N=1,D=10$-supersymmetries, the
bosonic sector has the same number of degrees of freedom as the
bosonic sector of an $N=2,D=4$ and $N=4,D=4$, respectively
\cite{2b}. Once the bosonic sectors are identified, we adopt an
$N=1,D=4$-superfield formulation to write down the gauge potential and the
Lorentz-violating background supermultiplets to finally set up their coupling in
terms of  $N=2$ and $N=4$ actions realized in $N=1$-superspace.
The result is projected out in component fields and we end up with the
complete actions that realize the extended supersymmetric version of the
Abelian Chern-Simons Lorentz-violating term.

The general organization of our work is as follows: in Section 2,
we set some preliminaries for the presentation of the $N=2$
Abelian gauge model in terms of $N=1$ superspace and superfields.
In Section 3, we focus on the task of carrying out the $N=2$
extension of the Lorentz-violating Chern-Simons term. Next, we go
one step further and reassess the discussion of Section 3 for the
case of a (maximally) $N=4$-extended gauge theory. This is the
content of Section 4. Finally, in Section 5, we present our
Concluding Remarks and Comments. An Appendix follows, where we
collect the relevant conventions to perform the $N=1$-superfield
manipulations.

\section{$N=2$-SUSY Abelian gauge model: basic ideas}

The $N=2$ supersymmetric generalization of the Abelian gauge model
can be built up by using the superfield formalism in an $N=1$
superspace parametrized by the coordinates  $(x^{\mu },\theta
^{a},\bar{\theta}_{\dot{a}})$ \cite{2}. The bosonic sector of the
gauge action can be obtained by means of a dimensional reduction
from $D=6$ to $D=4$ \cite{2c,2d}. The Maxwell Lagrangian in 6
dimensions is
\begin{equation}
{\cal L}=-\frac{1}{4}F_{\hat{\mu}\hat{\nu}}F^{\hat{\mu}\hat{\nu}},
\label{1}
\end{equation}
where $\hat{\mu}=0,1,2,3,4,5.$ The connection $A_{\hat{\mu}}$ can
be
parametrized as $A_{\hat{\mu}}=(A_{\mu },\varphi _{1},\varphi _{2}),$ where $%
\mu =0,1,2,3$. Notice that we keep the 6 components in 4
dimensions. By adopting as an ansatz
the fact that the fields have no dependence on the coordinates $%
x^{4},\,x^{5} $, we obtain the $D=4$ Lagrangian
\begin{equation}
{\cal L}=-\frac{1}{4}F_{\mu \nu }F^{\mu \nu }+\partial _{\mu
}\varphi
\partial ^{\mu }\varphi ^{*},  \label{2}
\end{equation}
where we define $\varphi =\varphi _{1}+i\varphi _{2}.$

This is the bosonic sector of the $N=2$ extended supersymmetric
action. The supersymmetrization of the theory above is achieved by
combining superfields of the $N=1$ superspace as supermultiplets
that accommodate  the ordinary fields and their superpartners. The superfields
that accomplish the task of accommodating the usual fields and their respective
superpartners are a scalar, $\Phi ,$ and vector superfield, $V,$ of $N=1-D=4$%
-superspace, which play the role of the vector multiplet $(\Phi ,V)$ for
$N=2-D=4$.

The vector superfield $V$ in the WZ-gauge is written as:
\begin{equation}
V=\theta \sigma ^{\mu }\bar{\theta}A_{\mu }+\theta ^{2}\bar{\theta}\bar{%
\lambda}+\bar{\theta}^{2}\theta \lambda +\theta
^{2}\bar{\theta}^{2}D, \label{3}
\end{equation}
which fulfills the reality constraint, $V=V^{\dagger }.$\ The
scalar superfield is written as
\begin{equation}
\Phi =\varphi +i\theta \sigma ^{\mu }\bar{\theta}\partial _{\mu }\varphi -%
\frac{1}{4}\theta ^{2}\bar{\theta}^{2}\Box \varphi +\sqrt{2}\theta \psi +%
\frac{i}{\sqrt{2}}\theta ^{2}\partial _{\mu }\psi \sigma ^{\mu }\bar{\theta}%
+\theta ^{2}f,  \label{4}
\end{equation}
\begin{equation}
\bar{\Phi}=\varphi ^{*}-i\theta \sigma ^{\mu }\bar{\theta}\partial
_{\mu
}\varphi ^{*}-\frac{1}{4}\theta ^{2}\bar{\theta}^{2}\Box \varphi ^{*}+\sqrt{2%
}\bar{\theta}\bar{\psi}+\frac{i}{\sqrt{2}}\bar{\theta}^{2}\theta
\sigma ^{\mu }\partial _{\mu }\bar{\psi}+\bar{\theta}^{2}f^{*},
\end{equation}
which obeys the chiral condition: $\bar{D}\Phi =D\bar{\Phi}=0.$

The $N=1$ scalar multiplet $\Phi $ is composed by spins
$(\frac{1}{2},0)$ and the vector multiplet encompasses the spins
$(1,\frac{1}{2}).$ Then, the vector
hypermultiplet $(\Phi ,V)$ in $N=2$ is composed by the spins $(1,\frac{1}{2}\,,%
\frac{1}{2},0,0)$ \cite{2}.

The $N=2$-supersymmetric extension of the Maxwell action contains the bosonic
gauge Lagrangian (\ref{2}) and is written as follows:
\begin{equation}
{\cal L}=\bar{\Phi}\Phi +W^{\alpha }W_{\alpha }\delta (\bar{\theta}^{2})+%
\bar{W}_{\dot{\alpha}}\bar{W}^{\dot{\alpha}}\delta (\theta ^{2}),
\label{5}
\end{equation}
where the Abelian field-strength superfield is given by:
\begin{equation}
W_{a}=-\frac{1}{4}\bar{D}^{2}D_{a}V_{WZ},\,\,\,\,\,\,\,\,\,\,\bar{W}_{\dot{a}%
}=-\frac{1}{4}D^{2}\bar{D}_{\dot{a}}V_{WZ},  \label{6}
\end{equation}
having the chirality condition: $\bar{D}W=D\bar{W}=0$ and $DW=\bar{D}\bar{W}%
. $

It is clear that the Lagrangian (\ref{5}) is invariant under $N=1$
supersymmetry transformation and it  also exhibits $N=2$ invariance.

\section{The Lorentz-violating term in the $N=2$ gauge model }

Now, we shall look for the $N=2$ supersymmetric version of the
Chern-Simons
Lorentz-breaking term. Using the fact that the bosonic sector for $N=2$ in $%
D=4$ is the same as the one for $N=1$ in $D=6$, we write the Chern-Simons term for $%
D=6$ and perform the dimensional reducion to $D=4$. The $D=4$
Chern-Simons term originally proposed by \cite{1} is
\begin{equation}
{\cal L}_{br}=\varepsilon ^{\mu \nu \kappa \lambda }A_{\mu
}\partial _{\nu }A_{\kappa }T_{\lambda }.  \label{7}
\end{equation}
We adopt for $D=6$ the Chern-Simons term in the form
\begin{equation}
{\cal L}_{br}=\varepsilon ^{\hat{\mu}\hat{\nu}\hat{\kappa}\hat{\lambda}\hat{%
\rho}\hat{\sigma}}A_{\hat{\mu}}\partial _{\hat{\nu}}A_{\hat{\kappa}}T_{\hat{%
\lambda}\hat{\rho}\hat{\sigma}}.  \label{7a}
\end{equation}
The background tensor $T_{\hat{\lambda}\hat{\rho}\hat{\sigma}}$
has 20 components, and we may rewrite  it as
\[
T_{\hat{\lambda}\hat{\rho}\hat{\sigma}}\equiv (R_{\rho \sigma
};S_{\rho \sigma };\partial _{\mu }v;\partial _{\mu }u),
\]
where $\hat{\mu}=\mu ,4,5$, and we consider there is no dependence
of the
fields on the $x^{4},x^{5}$ coordinates. The fields $R_{\rho \sigma }$ and $%
S_{\rho \sigma }$ have 6 components each one, and the other 8
components are redefined as 2 vectors that we write as a gradient
of the scalars fields $v$ and $u$. Then, the number of components
is reduced to 14.

As shown in the previous section, we also redefine the gauge field $A_{\hat{\mu}%
}\equiv (A_{\mu };\varphi _{1};\varphi _{2}).$ It is clear that
$\varepsilon
^{\hat{\mu}\hat{\nu}\hat{\kappa}\hat{\lambda}\hat{\rho}\hat{\sigma}}A_{\hat{%
\mu}}A_{\hat{\kappa}}\partial _{\hat{\nu}}T_{\hat{\lambda}\hat{\rho}\hat{%
\sigma}}=0,$ so we obtain, upon integration by parts, the Lagrangian as
follows:
\begin{eqnarray}
{\cal L}_{br} &=&-\frac{1}{4}\varepsilon ^{\mu \nu \kappa \lambda
}F_{\mu \nu }A_{\kappa }\partial _{\lambda
}v+\frac{1}{4}\varepsilon ^{\mu \nu
\kappa \lambda }F_{\mu \nu }\varphi _{1}R_{\kappa \lambda }+\frac{1}{4}%
\varepsilon ^{\mu \nu \kappa \lambda }F_{\mu \nu }\varphi
_{2}S_{\kappa
\lambda }  \label{8} \\
&&+\frac{1}{2}\varphi _{1}\partial _{\nu }\varphi _{2}\partial ^{\nu }u-%
\frac{1}{2}\varphi _{2}\partial _{\nu }\varphi _{1}\partial ^{\nu
}u. \nonumber
\end{eqnarray}
In order to carry out the supersymmetrization of the Lagrangian
(\ref{8}) by using a superspace formalism, it is advisable to define some
complex field combinations that are found in the superfields we deal with. We define these
bosonic fields as
\begin{eqnarray}
B_{\mu \nu } &=&S_{\mu \nu }-i\tilde{S}_{\mu \nu },  \nonumber \\
H_{\mu \nu } &=&R_{\mu \nu }-i\tilde{R}_{\mu \nu }  \nonumber \\
\varphi &=&\varphi _{1}+i\varphi _{2},  \label{8b} \\
r &=&t+iu,  \nonumber \\
s &=&w+iv,  \nonumber
\end{eqnarray}

Notice that we have introduced the new real scalar fields $t$ and
$w$ that are bosonic fields but do not appear in the bosonic
Lagrangian (\ref{8}). These fields will be necessary in the
supersymmetric version to maintain the balance between the bosonic
and fermionic degrees of freedom present in the scalar
superfields  defined with complex scalar fields. Each tensor
field, $R_{\mu \nu }$ and $S_{\mu \nu },$ appears as the real part
of the complex tensor field whose imaginary parts are given in
terms of their dual fields, as we see in (\ref{8b}) and can be
found in \cite{5}.

The superfields for the gauge sector have been defined above.
Thus, we take superfields which contain the fundamental fields of
the background sector plus their supersymmetric partners. These
superfields are $N=1$-multiplets that form an
$N=2$-hypermultiplet, $(S,R,\Sigma _{a},\Omega _{a}).$ The scalar
superfields that accommodate $s,\,s^{*},\,r$ and $r^{*}$ are,
respectively:
\begin{equation}
S=s+i\theta \sigma ^{\mu }\bar{\theta}\partial _{\mu
}s-\frac{1}{4}\theta ^{2}\bar{\theta}^{2}\Box s+\sqrt{2}\theta \xi
+\frac{i}{\sqrt{2}}\theta ^{2}\partial _{\mu }\xi \sigma ^{\mu
}\bar{\theta}+\theta ^{2}h,  \label{10}
\end{equation}
\begin{equation}
\bar{S}=s^{*}-i\theta \sigma ^{\mu }\bar{\theta}\partial _{\mu }s^{*}-\frac{1%
}{4}\theta ^{2}\bar{\theta}^{2}\Box s^{*}+\sqrt{2}\bar{\theta}\bar{\xi}+%
\frac{i}{\sqrt{2}}\bar{\theta}^{2}\theta \sigma ^{\mu }\partial _{\mu }\bar{%
\xi}+\bar{\theta}^{2}h^{*},  \label{10a}
\end{equation}
\begin{equation}
R=r+i\theta \sigma ^{\mu }\bar{\theta}\partial _{\mu
}r-\frac{1}{4}\theta ^{2}\bar{\theta}^{2}\Box r+\sqrt{2}\theta
\zeta +\frac{i}{\sqrt{2}}\theta ^{2}\partial _{\mu }\zeta \sigma
^{\mu }\bar{\theta}+\theta ^{2}g, \label{11}
\end{equation}
\begin{equation}
\bar{R}=r^{*}-i\theta \sigma ^{\mu }\bar{\theta}\partial _{\mu }r^{*}-\frac{1%
}{4}\theta ^{2}\bar{\theta}^{2}\Box r^{*}+\sqrt{2}\bar{\theta}\bar{\zeta}+%
\frac{i}{\sqrt{2}}\bar{\theta}^{2}\theta \sigma ^{\mu }\partial _{\mu }\bar{%
\zeta}+\bar{\theta}^{2}g^{*},
\end{equation}
which satisfy the chiral condition:
$\bar{D}S=D\bar{S}=\bar{D}R=D\bar{R}=0.$

The spinor superfields that contain $R_{\mu \nu },S_{\mu \nu }$
and their dual fields are written as
\begin{eqnarray}
\Sigma _{a} &=&\tau _{a}+\theta ^{b}(\varepsilon _{ba}\rho +\sigma
_{ba}^{\mu \nu }B_{\mu \nu })+\theta ^{2}F_{a}+i\theta \sigma ^{\mu }\bar{%
\theta}\partial _{\mu }\tau _{a}  \label{12} \\
&&+i\theta \sigma ^{\mu }\bar{\theta}\theta ^{b}\partial _{\mu
}(\varepsilon
_{ba}\rho +\sigma _{ba}^{\mu \nu }B_{\mu \nu })-\frac{1}{4}\theta ^{2}\bar{%
\theta}^{2}\square \tau _{a},  \nonumber
\end{eqnarray}
\begin{eqnarray}
\bar{\Sigma}_{\dot{a}} &=&\bar{\tau}_{\dot{a}}+\bar{\theta}_{\dot{b}%
}(-\varepsilon _{\,\,\dot{a}}^{\dot{b}}\rho ^{*}-\bar{\sigma}%
_{\,\,\,\,\,\,\,\,\,\,\,\dot{a}}^{\mu \nu \dot{b}}B_{\mu \nu }^{*})+\bar{%
\theta}^{2}\bar{F}_{\dot{a}}-i\theta \sigma ^{\mu
}\bar{\theta}\partial
_{\mu }\bar{\tau}_{\dot{a}} \\
&&-i\theta \sigma ^{\mu }\bar{\theta}\theta _{\dot{b}}\partial
_{\mu
}(-\varepsilon _{\,\,\,\dot{a}}^{\dot{b}}\rho ^{*}-\bar{\sigma}%
_{\,\,\,\,\,\,\,\,\,\,\,\dot{a}}^{\mu \nu \dot{b}}B_{\mu \nu }^{*})-\frac{1}{%
4}\theta ^{2}\bar{\theta}^{2}\square \bar{\tau}_{\dot{a}},
\nonumber
\end{eqnarray}
\begin{eqnarray}
\Omega _{a} &=&\chi _{a}+\theta ^{b}(\varepsilon _{ba}\phi +\sigma
_{ba}^{\mu \nu }H_{\mu \nu })+\theta ^{2}G_{a}+i\theta \sigma ^{\mu }\bar{%
\theta}\partial _{\mu }\chi _{a}  \label{12a} \\
&&+i\theta \sigma ^{\mu }\bar{\theta}\theta ^{b}\partial _{\mu
}(\varepsilon
_{ba}\phi +\sigma _{ba}^{\mu \nu }H_{\mu \nu })-\frac{1}{4}\theta ^{2}\bar{%
\theta}^{2}\square \chi _{a},  \nonumber
\end{eqnarray}
\begin{eqnarray}
\bar{\Omega}_{\dot{a}} &=&\bar{\chi}_{\dot{a}}+\bar{\theta}_{\dot{b}%
}(-\varepsilon _{\,\,\dot{a}}^{\dot{b}}\phi ^{*}-\bar{\sigma}%
_{\,\,\,\,\,\,\,\,\,\,\,\dot{a}}^{\mu \nu \dot{b}}H_{\mu \nu }^{*})+\bar{%
\theta}^{2}\bar{G}_{\dot{a}}-i\theta \sigma ^{\mu
}\bar{\theta}\partial
_{\mu }\bar{\chi}_{\dot{a}}  \label{12aa} \\
&&-i\theta \sigma ^{\mu }\bar{\theta}\theta _{\dot{b}}\partial
_{\mu
}(-\varepsilon _{\,\,\,\dot{a}}^{\dot{b}}\phi ^{*}-\bar{\sigma}%
_{\,\,\,\,\,\,\,\,\,\,\,\dot{a}}^{\mu \nu \dot{b}}H_{\mu \nu }^{*})-\frac{1}{%
4}\theta ^{2}\bar{\theta}^{2}\square \bar{\chi}_{\dot{a}},
\nonumber
\end{eqnarray}
that are also chiral $\bar{D}_{\dot{b}}\Sigma _{a}=D_{b}\bar{\Sigma}_{\dot{a}%
}=\bar{D}_{\dot{b}}\Omega _{a}=D_{b}\bar{\Omega}_{\dot{a}}=0.$ We
can notice that we have to introduce two extra background complex
scalar fields, $\rho $ and $\phi ,$ to match the bosonic and
fermionic degrees of freedom.

Now, we are interested in building up the supersymmetric action.
For that, it is useful to quote the mass dimensions of the
superfields previously given:
\begin{eqnarray*}
\lbrack \Phi ] &=&[\bar{\Phi}]=1,\,\,\,\,\,\,\,\,[V]=0,\,\,\,\,\,\,\,%
\,[W_{a}]=[\bar{W}_{\dot{a}}]=\frac{3}{2}, \\
\lbrack S] &=&[\bar{S}]=0,\,\,\,\,\,\,\,\,[\Sigma _{a}]=[\bar{\Sigma}_{\dot{a%
}}]=[\Omega
_{a}]=[\bar{\Omega}_{\dot{a}}]=1,\,\,\,\,\,\,\,\,[R]=[\bar{R}]=0.
\end{eqnarray*}

Based on the dimensionalities, and by analysing the bosonic
Lagrangian (\ref {8}), we propose the following supersymmetric
action, $S_{br}$:

\begin{eqnarray}
{\cal S}_{br} &=&\int d^{4}xd^{2}\theta d^{2}\bar{\theta}[\frac{1}{4}%
W^{a}(D_{a}V)S+\frac{1}{4}\bar{W}_{\dot{a}}(\bar{D}^{\dot{a}}V)\bar{S}+\frac{%
i}{4}\delta (\bar{\theta})W^{a}(\Phi +\bar{\Phi})\Sigma _{a}  \nonumber \\
&&-\frac{i}{4}\delta (\theta )\bar{W}_{\dot{a}}(\Phi +\bar{\Phi})\bar{\Sigma}%
^{\dot{a}}+\frac{1}{4}\delta (\bar{\theta})W^{a}(\Phi
-\bar{\Phi})\Omega _{a}
\label{13} \\
&&-\frac{1}{4}\delta (\theta )\bar{W}_{\dot{a}}(\Phi -\bar{\Phi})\bar{\Omega}%
^{\dot{a}}+\frac{1}{4}\Phi \bar{\Phi}(\bar{R}+R)],  \nonumber
\end{eqnarray}
which is invariant under the Abelian gauge transformations:
\begin{eqnarray}
\delta V &=&\Lambda -\bar{\Lambda} \\
\delta \Phi &=&\delta \bar{\Phi}=\delta S=\delta \bar{S}=\delta
R=\delta \bar{R}=\delta \Sigma _{a}=\delta
\bar{\Sigma}^{\dot{a}}=0.
\end{eqnarray}
In terms of superfields, we have two sectors:
\begin{eqnarray*}
Gauge \,\, sector &:&\,\,\,\,\,\{V,\Phi ,\bar{\Phi}\}
\\
Background \,\, sector\,\, &:&\,\,\,\,\,\{S,\bar{S}%
,\Omega _{a},\bar{\Omega}^{\dot{a}},\Sigma _{a},\bar{\Sigma}^{\dot{a}},R,%
\bar{R}\},
\end{eqnarray*}
and, in components, these two sectors have the field content
below:
\begin{eqnarray*}
Bosonic\,\, gauge\,\, sector &:&\,\,\,\,\,\{A_{\mu
},\varphi ,\varphi ^{*}\} \\
Fermionic\,\, gauge\,\, sector &:& \,\,\,\{\lambda
,\bar{\lambda},\psi
,\bar{\psi}\} \\
Bosonic\,\, background\,\, sector &:&\,\,\,\,\,%
\{s,s^{*},R_{\mu \nu },S_{\mu \nu },\rho ,\rho ^{*},\phi ,\phi
^{*},r,r^{*}\}
\\
Fermionic\,\, background\,\, sector &:&\,\,\,\,\,\{\xi ,\bar{\xi},\tau ,\bar{%
\tau},F,\bar{F},\chi ,\bar{\chi},G,\bar{G},\zeta ,\bar{\zeta}\}.
\end{eqnarray*}
We therefore observe that the action (\ref{13}) is manifestly
invariant
under $N=1$-supersymmetry. The component-field content of the $N=2$%
-supersymmetry is accommodated in the $N=1$-superfields given in equations (%
\ref{10})-(\ref{12aa}). Indeed, the action (\ref{13}) displays a
larger supersymmetry, $N=2,$ realised in terms of an $N=1$
-superspace formulation.

This Lagrangian in its component-field version reads as below:
\begin{eqnarray}
{\cal L}_{br} &=&+\frac{i}{8}\partial _{\mu }(s-s^{*})\varepsilon
^{\mu \kappa \lambda \nu }F_{\kappa \lambda }A_{\nu
}-\frac{1}{8}(s+s^{*})F_{\mu
\nu }F^{\mu \nu }+D^{2}(s+s^{*})  \nonumber \\
&&-\frac{1}{2}is\lambda \sigma ^{\mu }\partial _{\mu }\bar{\lambda}-\frac{1}{%
2}is^{*}\bar{\lambda}\bar{\sigma}^{\mu }\partial _{\mu }\lambda -\frac{1}{2%
\sqrt{2}}\lambda \sigma ^{\mu \nu }F_{\mu \nu }\xi +\frac{1}{2\sqrt{2}}\bar{%
\lambda}\bar{\sigma}^{\mu \nu }F_{\mu \nu }\bar{\xi}  \nonumber \\
&&+\frac{1}{4}\lambda \lambda h+\frac{1}{4}\bar{\lambda}\bar{\lambda}h^{*}-%
\frac{1}{\sqrt{2}}\lambda \xi
D-\frac{1}{\sqrt{2}}\bar{\lambda}\bar{\xi}D
\nonumber \\
&&\frac{1}{16}\varepsilon ^{\mu \nu \kappa \lambda }F_{\mu \nu
}(\varphi
+\varphi ^{*})(B_{\kappa \lambda }+B_{\kappa \lambda }^{*})+\frac{i}{8}%
F^{\mu \nu }(B_{\mu \nu }-B_{\mu \nu }^{*})(\varphi +\varphi ^{*})
\nonumber
\\
&&-\frac{i\sqrt{2}}{8}\tau \sigma ^{\mu \nu }\psi F_{\mu \nu }-\frac{i\sqrt{2%
}}{8}\bar{\tau}\bar{\sigma}^{\mu \nu }\bar{\psi}F_{\mu \nu
}+\frac{1}{4}\tau \sigma ^{\mu }\partial _{\mu
}\bar{\lambda}(\varphi +\varphi ^{*})  \nonumber
\\
&&-\frac{1}{4}\bar{\tau}\bar{\sigma}^{\mu }\partial _{\mu }\lambda
(\varphi +\varphi ^{*})+\frac{i\sqrt{2}}{4}\psi \sigma ^{\mu \nu
}B_{\mu \nu }\lambda
+\frac{i\sqrt{2}}{4}\bar{\psi}\bar{\sigma}^{\mu \nu }B_{\mu \nu }^{*}\bar{%
\lambda}  \nonumber \\
&&-\frac{i}{2}D(\varphi +\varphi ^{*})\rho
+\frac{i}{2}D^{*}(\varphi
+\varphi ^{*})\rho ^{*}  \nonumber \\
&&+\frac{i\sqrt{2}}{8}\lambda \psi \rho -\frac{i\sqrt{2}}{8}\bar{\lambda}%
\bar{\psi}\rho ^{*}-\frac{i\sqrt{2}}{4}D\psi \tau +\frac{i\sqrt{2}}{4}D^{*}%
\bar{\psi}\bar{\tau}  \label{14} \\
&&+\frac{i}{4}f\lambda \tau -\frac{i}{4}f^{*}\bar{\lambda}\bar{\tau}+\frac{i%
}{4}(\varphi +\varphi ^{*})\lambda F-\frac{i}{4}(\varphi +\varphi ^{*})\bar{%
\lambda}\bar{F}  \nonumber \\
&&-\frac{i}{16}\varepsilon ^{\mu \nu \kappa \lambda }F_{\mu \nu
}(\varphi
-\varphi ^{*})(H_{\kappa \lambda }+H_{\kappa \lambda }^{*})+\frac{1}{8}%
F^{\mu \nu }(H_{\kappa \lambda }-H_{\kappa \lambda }^{*})(\varphi
-\varphi
^{*})  \nonumber \\
&&-\frac{\sqrt{2}}{8}\chi \sigma ^{\mu \nu }\psi F_{\mu \nu }+\frac{\sqrt{2}%
}{8}\bar{\chi}\bar{\sigma}^{\mu \nu }\bar{\psi}F_{\mu \nu
}-\frac{i}{4}\chi \sigma ^{\mu }\partial _{\mu
}\bar{\lambda}(\varphi -\varphi ^{*})  \nonumber
\\
&&+\frac{i}{4}\bar{\chi}\bar{\sigma}^{\mu }\partial _{\mu }\lambda
(\varphi
-\varphi ^{*})+\frac{\sqrt{2}}{4}\psi \sigma ^{\mu \nu }H_{\mu \nu }\lambda -%
\frac{\sqrt{2}}{4}\bar{\psi}\bar{\sigma}^{\mu \nu }H_{\mu \nu }^{*}\bar{%
\lambda}  \nonumber \\
&&-\frac{1}{2}D(\varphi -\varphi ^{*})\phi
+\frac{1}{2}D^{*}(\varphi
-\varphi ^{*})\phi ^{*}  \nonumber \\
&&+\frac{\sqrt{2}}{8}\lambda \psi \phi +\frac{\sqrt{2}}{8}\bar{\lambda}\bar{%
\psi}\phi ^{*}-\frac{\sqrt{2}}{4}D\psi \chi -\frac{\sqrt{2}}{4}D^{*}\bar{\psi%
}\bar{\chi}  \nonumber \\
&&+\frac{1}{4}f\lambda \chi +\frac{1}{4}f^{*}\bar{\lambda}\bar{\chi}+\frac{1%
}{4}(\varphi -\varphi ^{*})\lambda G-\frac{1}{4}(\varphi -\varphi ^{*})\bar{%
\lambda}\bar{G}  \nonumber \\
&&+\frac{1}{8}\varphi \partial _{\mu }\varphi ^{*}\partial ^{\mu }(r-r^{*})-%
\frac{1}{8}\varphi ^{*}\partial _{\mu }\varphi \partial ^{\mu
}(r-r^{*})
\nonumber \\
&&+\frac{1}{4}\partial ^{\mu }\varphi \partial _{\mu }\varphi ^{*}(r+r^{*})-%
\frac{1}{8}\varphi \varphi ^{*}\square (r+r^{*})-\frac{i}{4}\psi
\sigma
^{\mu }\partial _{\mu }\bar{\psi}(r+r^{*})  \nonumber \\
&&+\frac{1}{4}ff^{*}(r+r^{*})-\frac{i}{4}\psi \sigma ^{\mu }\bar{\psi}%
\partial _{\mu }r^{*}  \nonumber \\
&&-\frac{i}{4}\varphi \zeta \sigma ^{\mu }\partial _{\mu }\bar{\psi}-\frac{i%
}{4}\varphi ^{*}\psi \sigma ^{\mu }\partial _{\mu }\bar{\zeta}-\frac{i}{4}%
\psi \sigma ^{\mu }\bar{\zeta}\partial _{\mu }\varphi ^{*}  \nonumber \\
&&+\frac{1}{4}\varphi f^{*}g+\frac{1}{4}f\varphi ^{*}g^{*}-\frac{1}{4}%
f^{*}\psi \zeta -\frac{1}{4}f\bar{\psi}\bar{\zeta}.  \nonumber
\end{eqnarray}
We point out the pieces corresponding to the bosonic action
(\ref{8}) in the complete component-field action above:
\begin{eqnarray*}
\frac{i}{8}\partial _{\mu }(s-s^{*})\varepsilon ^{\mu \kappa
\lambda \nu }F_{\kappa \lambda }A_{\nu }
&=&-\frac{1}{4}\varepsilon ^{\mu \nu \kappa
\lambda }F_{\mu \nu }A_{\kappa }\partial _{\lambda }v, \\
\frac{1}{16}\varepsilon ^{\mu \nu \kappa \lambda }F_{\mu \nu
}(\varphi
+\varphi ^{*})(B_{\kappa \lambda }+B_{\kappa \lambda }^{*}) &=&\frac{1}{4}%
\varepsilon ^{\mu \nu \kappa \lambda }F_{\mu \nu }\varphi
_{1}R_{\kappa
\lambda }, \\
-\frac{i}{16}\varepsilon ^{\mu \nu \kappa \lambda }F_{\mu \nu
}(\varphi
-\varphi ^{*})(H_{\kappa \lambda }+H_{\kappa \lambda }^{*}) &=&\frac{1}{4}%
\varepsilon ^{\mu \nu \kappa \lambda }F_{\mu \nu }\varphi
_{2}S_{\kappa
\lambda }, \\
\frac{1}{8}\varphi \partial _{\mu }\varphi ^{*}\partial ^{\mu }(r-r^{*})-%
\frac{1}{8}\varphi ^{*}\partial _{\mu }\varphi \partial ^{\mu }(r-r^{*}) &=&%
\frac{1}{2}\varphi _{1}\partial _{\nu }\varphi _{2}\partial ^{\nu }u-\frac{1%
}{2}\varphi _{2}\partial _{\nu }\varphi _{1}\partial ^{\nu }u.
\end{eqnarray*}
We can notice that this Lagrangian describes the bosonic sector
(\ref{8}) and its superpartners. We find here the $N=1$
supersymmetrization of the Chern-Simons term presented in
\cite{3}, where the first term is the same as proposed by
\cite{1}, considering the constant vector as the gradient of a
scalar. Since the gradient vector is a constant, we have that
$s=\alpha +\beta ^{\mu }x_{\mu }.$ We notice in our Lagrangian the presence of the
bosonic real scalar fields, $t=s+s^{*}$ and
$u=r+r^{*},$ and the complex scalar fields, $\rho $ and $\phi ,$
that do not appear in the bosonic Lagrangian (\ref{8}). These
scalar fields appear in the supersymmetric generalization in order
to keep the bosonic and fermionic degrees of freedom
in equal number. We can see that the bosonic fields $%
D,D^{*},f,f^{*},h,h^{*},g$ and $g^{*}$ play all the role of
auxiliary fields. The bosonic fields $s,s^{*},R_{\mu \nu },S_{\mu
\nu },\rho ,\rho
^{*},\phi ,\phi ^{*},r,r^{*}$ and the fermionic fields $\xi ,\bar{\xi},\tau ,%
\bar{\tau},F,\bar{F},\chi ,\bar{\chi},G,\bar{G},\zeta
,\bar{\zeta}$ work as background fields breaking the Lorentz
invariance.

\section{$N=4$-Supersymmetric Extension of the Lorentz-Violating Action}

The $N=4$ supersymmetric generalization of the Abelian gauge model
in $D=4$
can be built up in an $N=1$ superspace background, with coordinates $%
(x^{\mu },\theta ^{a},\bar{\theta}_{\dot{a}})$ \cite{2}. The
bosonic sector
of the gauge action can be obtained upon a dimensional reduction from $%
D=10$ to $D=4$. The Maxwell Lagrangian in 10 dimensions is
\begin{equation}
{\cal L}=-\frac{1}{4}F_{\hat{\mu}\hat{\nu}}F^{\hat{\mu}\hat{\nu}},
\label{15}
\end{equation}
where $\hat{\mu}=\mu ,4,5,6,7,8,9$ and $\mu =0,1,2,3.$ The connection $A_{%
\hat{\mu}}$ can be parametrised as $A_{\hat{\mu}}=(A_{\mu
},\varphi ^{I},\,I=1,2,3,4,5,6)$. Notice that we keep the 10 field
components in 4 dimensions. Adopting the fact that the fields have no
dependence on the coordinates
$x^{4},\,x^{5},x^{6},\,x^{7},x^{8},\,x^{9}$, we obtain the
Lagrangian
\begin{equation}
{\cal L}=-\frac{1}{4}F_{\mu \nu }F^{\mu \nu }+\partial _{\mu
}\varphi ^{I}\partial ^{\mu }\varphi ^{I}.  \label{16}
\end{equation}

This is the bosonic sector of the $N=4$-extended supersymmetric
action. The supersymmetrization of the theory above is
accomplished by defining superfields in the $N=1$ superspace as
multiplets containing the fields and their superpartners. The
superfields that contain these bosonic fields and their
superpartners are 6 chiral scalars, $\Phi ^{I},$ and vector
superfield, $V,$ of $N=1$-superspace; put together, $(\Phi
^{I},V),$ they form the gauge multiplet of $N=4$-supersymmetry.

The vector superfield $V$ as defined in (\ref{3}) fulfills the
reality constraint, $V=V^{\dagger }.$\ The scalar superfield is
written as
\begin{equation}
\Phi ^{I}=\varphi ^{I}+i\theta \sigma ^{\mu }\bar{\theta}\partial
_{\mu
}\varphi ^{I}-\frac{1}{4}\theta ^{2}\bar{\theta}^{2}\Box \varphi ^{I}+\sqrt{2%
}\theta \psi ^{I}+\frac{i}{\sqrt{2}}\theta ^{2}\partial _{\mu
}\psi ^{I}\sigma ^{\mu }\bar{\theta}+\theta ^{2}f^{I},  \label{18}
\end{equation}
\begin{equation}
\bar{\Phi}^{I}=\varphi ^{*I}-i\theta \sigma ^{\mu
}\bar{\theta}\partial _{\mu }\varphi ^{*I}-\frac{1}{4}\theta
^{2}\bar{\theta}^{2}\Box \varphi
^{*I}+\sqrt{2}\bar{\theta}\bar{\psi}^{I}+\frac{i}{\sqrt{2}}\bar{\theta}%
^{2}\theta \sigma ^{\mu }\partial _{\mu }\bar{\psi}^{I}+\bar{\theta}%
^{2}f^{*I},  \label{18b}
\end{equation}
which obeys the chirality condition: $\bar{D}\Phi
^{I}=D\bar{\Phi}^{I}=0.$

The $N=4$-supersymmetric extension of the gauge Lagrangian
(\ref{16}) is
\begin{equation}
{\cal L}=\bar{\Phi}^{I}\Phi ^{I}+W^{\alpha }W_{\alpha }\delta (\bar{\theta}%
^{2})+\bar{W}_{\dot{\alpha}}\bar{W}^{\dot{\alpha}}\delta (\theta
^{2}), \label{19}
\end{equation}
where the Abelian field-strength superfield was defined in
(\ref{6}).

It is clear that the Lagrangian (\ref{19}) is invariant under $N=1$%
-supersymmetry transformation and it has also $N=4$ invariance.

Now, we shall look for $N=4$ supersymmetric version of the
Chern-Simons Lorentz breaking term in $D=4$. Similarly to what we have done
for $N=2,$ we shall use the fact that the bosonic sector for $N=4$
in $D=4$ is the same for the $N=1$ in $D=10$ \cite{2}. We write
the Chern-Simons term for $D=10$ and perform its dimensional reduction
to $D=4$. We propose for $D=10$ the Chern-Simons term in the form
\begin{equation}
{\cal L}_{br}=\varepsilon ^{\hat{\mu}\hat{\nu}\hat{\kappa}\hat{\lambda}\hat{%
\rho}\hat{\sigma}\hat{\delta}\hat{\tau}\hat{\beta}\hat{\gamma}}A_{\hat{\mu}%
}\partial _{\hat{\nu}}A_{\hat{\kappa}}T_{\hat{\lambda}\hat{\rho}\hat{\sigma}%
\hat{\delta}\hat{\tau}\hat{\beta}\hat{\gamma}}.  \label{21}
\end{equation}
The background tensor $T_{\hat{\lambda}\hat{\rho}\hat{\sigma}}$
has 120 components, but we can redefine it as
\[
T_{\hat{\lambda}\hat{\rho}\hat{\sigma}}\equiv (R_{\rho \sigma
}^{I};\partial _{\mu }v;\partial _{\mu }u^{IJ}\,),
\]
where $\hat{\mu}=\mu ,4,5,6,7,8,9$ is the space-time index and $%
I,J=1,2,3,4,5,6$ is an internal index. We consider that there is no
dependence of the fields on the
$x^{4},x^{5},x^{6},x^{7},x^{8},x^{9}$ coordinates. Then, we have 6
anti-symmetric tensor fields $R_{\rho \sigma }^{I}$ with 6
components each one and 15 vectors written as gradients of 15
scalars represented by the anti-symmetric index $I,J.$ Then, the
number of components is reduced to 52.

Next, we need to redefine the gauge field as $A_{\hat{\mu}}\equiv
(A_{\mu };\varphi ^{I}\,,\,I=1,2,3,4,5,6)$ where $\varphi ^{I}$ is
real scalar
fields. Observing that $\varepsilon ^{\hat{\mu}\hat{\nu}\hat{\kappa}\hat{%
\lambda}\hat{\rho}\hat{\sigma}\hat{\delta}\hat{\tau}\hat{\beta}\hat{\gamma}%
}A_{\hat{\mu}}A_{\hat{\kappa}}\partial _{\hat{\nu}}T_{\hat{\lambda}\hat{\rho}%
\hat{\sigma}\hat{\delta}\hat{\tau}\hat{\beta}\hat{\gamma}}=0,$ we
obtain, integrating by parts, the Lagrangian as follows:
\begin{equation}
{\cal L}_{br}=-\frac{1}{4}\varepsilon ^{\mu \nu \kappa \lambda
}F_{\mu \nu }A_{\kappa }\partial _{\lambda
}v+\frac{1}{4}\varepsilon ^{\mu \nu \kappa \lambda }F_{\mu \nu
}\varphi ^{I}R_{\kappa \lambda }^{I}+\frac{1}{2}\varphi
^{I}\partial _{\nu }\varphi ^{J}\partial ^{\nu }u^{IJ}.
\label{22}
\end{equation}
As in the case of $N=2,$ we have to define some complex field combinations
that can be found in superfields. We define these bosonic fields
as
\begin{eqnarray}
B_{\mu \nu }^{I} &=&R_{\mu \nu }^{I}-i\tilde{R}_{\mu \nu }^{I},  \nonumber \\
\hat{\varphi}^{I} &=&\varphi ^{I}+i\beta ^{I},  \label{23} \\
r^{IJ} &=&t^{IJ}+iu^{IJ},  \nonumber \\
s &=&w+iv.  \nonumber
\end{eqnarray}

Notice that we have introduced the new real scalar fields $\beta
^{I},t^{IJ},w$ which are not present in the
bosonic Lagrangian (\ref{22}). As already pointed out, this has to be done
the supersymmetric version to maintain the same number of degree
for the matching of the bosonic and fermionic sectors of the scalar
superfields defined in terms of complex scalar fields. Each tensor
field, $R_{\mu \nu }^{I},$ appears as the real part of the complex
tensor field whose imaginary parts are given in terms of their
dual fields.

We now take the superfields which contain the fundamental fields of
the background sector and that accommodate their supersymmetric partners. These
superfields are $N=1$-multiplets that combine to form an
$N=4$-hypermultiplet, $(S,R^{IJ},\Sigma _{a}^{I}).$ The scalar
superfields that accommodate $s,s^{*},r^{*IJ}$ and $r^{IJ}$ are,
respectively, $S$ and $\bar{S}$ (as in eqs. (\ref{10}) and (\ref{10a})) and $%
R^{IJ}$ and $\bar{R}^{IJ}$ as cast below:
\begin{equation}
R^{IJ}=r^{IJ}+i\theta \sigma ^{\mu }\bar{\theta}\partial _{\mu }r^{IJ}-\frac{%
1}{4}\theta ^{2}\bar{\theta}^{2}\Box r^{IJ}+\sqrt{2}\theta \zeta ^{IJ}+\frac{%
i}{\sqrt{2}}\theta ^{2}\partial _{\mu }\zeta ^{IJ}\sigma ^{\mu }\bar{\theta}%
+\theta ^{2}g^{IJ},  \label{26}
\end{equation}
\begin{equation}
\bar{R}^{IJ}=r^{*IJ}-i\theta \sigma ^{\mu }\bar{\theta}\partial
_{\mu
}r^{*IJ}-\frac{1}{4}\theta ^{2}\bar{\theta}^{2}\Box r^{*IJ}+\sqrt{2}\bar{%
\theta}\bar{\zeta}^{IJ}+\frac{i}{\sqrt{2}}\bar{\theta}^{2}\theta
\sigma ^{\mu }\partial _{\mu
}\bar{\zeta}^{IJ}+\bar{\theta}^{2}g^{*IJ},
\end{equation}
which satisfy the chiral condition: $\bar{D}S=D\bar{S}=\bar{D}R^{IJ}=D\bar{R}%
^{IJ}=0.$

The spinor superfields that contain $R_{\mu \nu }^{I}$ and their
respective dual fields are written as
\begin{eqnarray}
\Sigma _{a}^{I} &=&\tau _{a}^{I}+\theta ^{b}(\varepsilon _{ba}\rho
^{I}+\sigma _{ba}^{\mu \nu }B_{\mu \nu }^{I})+\theta
^{2}F_{a}^{I}+i\theta
\sigma ^{\mu }\bar{\theta}\partial _{\mu }\tau _{a}^{I} \\
&&+i\theta \sigma ^{\mu }\bar{\theta}\theta ^{b}\partial _{\mu
}(\varepsilon _{ba}\rho ^{I}+\sigma _{ba}^{\mu \nu }B_{\mu \nu
}^{I})-\frac{1}{4}\theta ^{2}\bar{\theta}^{2}\square \tau
_{a}^{I},  \nonumber
\end{eqnarray}
\begin{eqnarray}
\bar{\Sigma}_{\dot{a}}^{I} &=&\bar{\tau}_{\dot{a}}^{I}+\bar{\theta}_{\dot{b}%
}(-\varepsilon _{\,\,\dot{a}}^{\dot{b}}\rho ^{*I}-\bar{\sigma}%
_{\,\,\,\,\,\,\,\,\,\,\,\dot{a}}^{\mu \nu \dot{b}}B_{\mu \nu }^{*I})+\bar{%
\theta}^{2}\bar{F}_{\dot{a}}^{I}-i\theta \sigma ^{\mu
}\bar{\theta}\partial
_{\mu }\bar{\tau}_{\dot{a}}^{I} \\
&&-i\theta \sigma ^{\mu }\bar{\theta}\theta _{\dot{b}}\partial
_{\mu
}(-\varepsilon _{\,\,\,\dot{a}}^{\dot{b}}\rho ^{*I}-\bar{\sigma}%
_{\,\,\,\,\,\,\,\,\,\,\,\dot{a}}^{\mu \nu \dot{b}}B_{\mu \nu }^{*I})-\frac{1%
}{4}\theta ^{2}\bar{\theta}^{2}\square \bar{\tau}_{\dot{a}}^{I},
\nonumber
\end{eqnarray}
that are also chiral $\bar{D}_{\dot{b}}\Sigma _{a}^{I}=D_{b}\bar{\Sigma}_{%
\dot{a}}^{I}=0.$ We can notice that in spinor superfields we have
to introduce six extra background complex scalar fields, $\rho
^{I}$, to match the bosonic and fermionic degrees of freedom.

Based on dimensional analysis arguments for the bosonic sector, as
it has been done for the $N=2$ case, and noticing that some
superfields now have internal symmetry index, we propose the
following $N=4$ supersymmetric action:

\begin{eqnarray}
{\cal S}_{br} &=&\int d^{4}xd^{2}\theta d^{2}\bar{\theta}[\frac{1}{4}%
W^{a}(D_{a}V)S+\frac{1}{4}\bar{W}_{\dot{a}}(\bar{D}^{\dot{a}}V)\bar{S}+\frac{%
i}{4}\delta (\bar{\theta})W^{a}(\Phi ^{I}+\bar{\Phi}^{I})\Sigma
_{a}^{I}
\label{40} \\
&&-\frac{i}{4}\delta (\theta )\bar{W}_{\dot{a}}(\Phi ^{I}+\bar{\Phi}^{I})%
\bar{\Sigma}^{\dot{a}I}+\frac{1}{4}\Phi ^{I}\bar{\Phi}^{J}(R^{IJ}-\bar{R}%
^{IJ})],  \nonumber
\end{eqnarray}
which is invariant under gauge transformations:
\begin{eqnarray}
\delta V &=&\Lambda -\bar{\Lambda} \\
\delta \Phi ^{I} &=&\delta \bar{\Phi}^{I}=\delta S=\delta
\bar{S}=\delta
R^{IJ}=\delta \bar{R}^{IJ}=\delta \Sigma _{a}^{I}=\delta \bar{\Sigma}^{\dot{a%
}I}=0.
\end{eqnarray}
In terms of superfields, we have two sectors:
\begin{eqnarray*}
Gauge\,\, Sector &:&\,\,\,\,\,\{V,\Phi ^{I}\} \\
Background\,\, Sector \,\, &:&\,\,\,\,\,\{\Sigma
_{a}^{I},\bar{\Sigma}^{\dot{a}I},S,\bar{S},R^{IJ},\bar{R}^{IJ}\},
\end{eqnarray*}
and, in components these two sectors encompass the fields cast
below:
\begin{eqnarray*}
Bosonic\,\, gauge\,\, Sector &:&\,\,\,\,\,\{A_{\mu
},\varphi ^{I},\varphi ^{*I}\} \\
Fermionic\,\, gauge\,\, Sector &:&\,\,\,\{\lambda
,\bar{\lambda},\psi
^{I},\bar{\psi}^{I}\} \\
Bosonic\,\, background\,\, Sector &:&\,\,\,\,\,%
\{s,s^{*},R_{\mu \nu }^{I},\rho ^{I},\rho ^{*I},r^{IJ},r^{*IJ}\} \\
Fermionic\,\, background\,\, Sector &:&\,\,\,\,\,\{\xi ,\bar{\xi},\tau ^{I},%
\bar{\tau}^{I},F^{I},\bar{F}^{I},\zeta ^{IJ},\bar{\zeta}^{IJ}\}.
\end{eqnarray*}
We can observe that the action (\ref{40}) is invariant under $N=1$%
-supersymmetry and there is a larger symmetry, the
$N=4$-supersymmetry as well.

This $N=4$ Lagrangian in its component-field version reads as
follows:
\begin{eqnarray}
{\cal L}_{br} &=&+\frac{i}{8}\partial _{\mu }(s-s^{*})\varepsilon
^{\mu \kappa \lambda \nu }F_{\kappa \lambda }A_{\nu
}-\frac{1}{8}(s+s^{*})F_{\mu
\nu }F^{\mu \nu }+D^{2}(s+s^{*})  \nonumber \\
&&-\frac{1}{2}is\lambda \sigma ^{\mu }\partial _{\mu }\bar{\lambda}-\frac{1}{%
2}is^{*}\bar{\lambda}\bar{\sigma}^{\mu }\partial _{\mu }\lambda -\frac{1}{2%
\sqrt{2}}\lambda \sigma ^{\mu \nu }F_{\mu \nu }\xi +\frac{1}{2\sqrt{2}}\bar{%
\lambda}\bar{\sigma}^{\mu \nu }F_{\mu \nu }\bar{\xi}  \nonumber \\
&&+\frac{1}{4}\lambda \lambda h+\frac{1}{4}\bar{\lambda}\bar{\lambda}h^{*}-%
\frac{1}{\sqrt{2}}\lambda \xi
D-\frac{1}{\sqrt{2}}\bar{\lambda}\bar{\xi}D
\nonumber \\
&&\frac{1}{16}\varepsilon ^{\mu \nu \kappa \lambda }F_{\mu \nu
}(\varphi
^{I}+\varphi ^{*I})(B_{\kappa \lambda }^{I}+B_{\kappa \lambda }^{*I})+\frac{i%
}{8}F^{\mu \nu }(B_{\mu \nu }^{I}-B_{\mu \nu }^{*I})(\varphi
^{I}+\varphi
^{*I})  \nonumber \\
&&-\frac{i\sqrt{2}}{8}\tau ^{I}\sigma ^{\mu \nu }\psi ^{I}F_{\mu \nu }-\frac{%
i\sqrt{2}}{8}\bar{\tau}^{I}\bar{\sigma}^{\mu \nu }\bar{\psi}^{I}F_{\mu \nu }+%
\frac{1}{4}\tau ^{I}\sigma ^{\mu }\partial _{\mu
}\bar{\lambda}(\varphi
^{I}+\varphi ^{*I})  \nonumber \\
&&-\frac{1}{4}\bar{\tau}^{I}\bar{\sigma}^{\mu }\partial _{\mu
}\lambda (\varphi ^{I}+\varphi ^{*I})+\frac{i\sqrt{2}}{4}\psi
^{I}\sigma ^{\mu \nu }B_{\mu \nu }^{I}\lambda
+\frac{i\sqrt{2}}{4}\bar{\psi}^{I}\bar{\sigma}^{\mu
\nu }B_{\mu \nu }^{*}\bar{\lambda}^{I}  \nonumber \\
&&-\frac{i}{2}D(\varphi ^{I}+\varphi ^{*I})\rho ^{I}+\frac{i}{2}%
D^{*}(\varphi ^{I}+\varphi ^{*I})\rho ^{*I}  \nonumber \\
&&+\frac{i\sqrt{2}}{8}\lambda \psi ^{I}\rho ^{I}-\frac{i\sqrt{2}}{8}\bar{%
\lambda}\bar{\psi}^{I}\rho ^{*I}-\frac{i\sqrt{2}}{4}D\psi ^{I}\tau ^{I}+%
\frac{i\sqrt{2}}{4}D^{*}\bar{\psi}^{I}\bar{\tau}^{I}  \label{30} \\
&&+\frac{i}{4}f^{I}\lambda \tau ^{I}-\frac{i}{4}f^{*I}\bar{\lambda}\bar{\tau}%
^{I}+\frac{i}{4}(\varphi ^{I}+\varphi ^{*I})\lambda F^{I}-\frac{i}{4}%
(\varphi ^{I}+\varphi ^{*I})\bar{\lambda}\bar{F}^{I}  \nonumber \\
&&+\frac{1}{8}\varphi ^{I}\partial _{\mu }\varphi ^{*J}\partial
^{\mu }(r^{IJ}+r^{*IJ})-\frac{1}{8}\varphi ^{*J}\partial _{\mu
}\varphi
^{I}\partial ^{\mu }(r^{IJ}+r^{*IJ})  \nonumber \\
&&+\frac{1}{4}\partial ^{\mu }\varphi ^{I}\partial _{\mu }\varphi
^{*J}(r^{IJ}-r^{*IJ})-\frac{1}{8}\varphi ^{I}\varphi ^{*J}\square
(r^{IJ}-r^{*IJ})-\frac{i}{4}\psi ^{I}\sigma ^{\mu }\partial _{\mu }\bar{\psi}%
^{J}(r^{IJ}-r^{*IJ})  \nonumber \\
&&+\frac{1}{4}f^{I}f^{*J}(r^{IJ}-r^{*IJ})+\frac{i}{4}\psi ^{I}\sigma ^{\mu }%
\bar{\psi}^{J}\partial _{\mu }r^{*IJ}  \nonumber \\
&&-\frac{i}{4}\varphi ^{I}\zeta ^{IJ}\sigma ^{\mu }\partial _{\mu }\bar{\psi}%
^{J}+\frac{i}{4}\varphi ^{*J}\psi ^{I}\sigma ^{\mu }\partial _{\mu }\bar{%
\zeta}^{IJ}+\frac{i}{4}\psi ^{I}\sigma ^{\mu
}\bar{\zeta}^{IJ}\partial _{\mu
}\varphi ^{*J}  \nonumber \\
&&+\frac{1}{4}\varphi ^{I}f^{*J}g^{IJ}-\frac{1}{4}f^{I}\varphi ^{*J}g^{*IJ}-%
\frac{1}{4}f^{*J}\psi ^{I}\zeta ^{IJ}+\frac{1}{4}f^{I}\bar{\psi}^{J}\bar{%
\zeta}^{IJ}.  \nonumber
\end{eqnarray}
We can ascertain the presence of the bosonic sector (\ref{22})
by means of the terms below:
\begin{eqnarray*}
\frac{i}{8}\partial _{\mu }(s-s^{*})\varepsilon ^{\mu \kappa
\lambda \nu }F_{\kappa \lambda }A_{\nu }
&=&-\frac{1}{4}\varepsilon ^{\mu \nu \kappa
\lambda }F_{\mu \nu }A_{\kappa }\partial _{\lambda }v, \\
\frac{1}{16}\varepsilon ^{\mu \nu \kappa \lambda }F_{\mu \nu
}(\varphi
^{I}+\varphi ^{*I})(B_{\kappa \lambda }^{I}+B_{\kappa \lambda }^{*I}) &=&%
\frac{1}{4}\varepsilon ^{\mu \nu \kappa \lambda }F_{\mu \nu
}\varphi
^{I}R_{\kappa \lambda }^{I}, \\
\frac{1}{8}\varphi ^{I}\partial _{\mu }\varphi ^{*J}\partial ^{\mu
}(r^{IJ}+r^{*IJ})-\frac{1}{8}\varphi ^{*J}\partial _{\mu }\varphi
^{I}\partial ^{\mu }(r^{IJ}+r^{*IJ}) &=&\frac{1}{2}(\varphi
^{I}\partial _{\nu }\varphi ^{J}+\beta ^{I}\partial _{\nu }\beta
^{J})\partial ^{\nu }u^{IJ}.
\end{eqnarray*}
We can notice that this Lagrangian fairly accomodates the $N=4$
bosonic sector (\ref{22}). We re-obtain here the $N=1$ and $N=2$
supersymmetrisation of the Chern-Simons term presented in
ref.\cite{3} and in (\ref{14}), respectively. We notice that $N=4$
Lagrangian is similar to $N=2$ but now existing an internal index
in same fields. The fields $\beta ^{I},\,t,\,\,u^{IJ}$ and $\,\rho
^{IJ},$ that do not appear in the bosonic Lagrangian (\ref{22}),
were introduced in order to keep the bosonic and fermionic degrees
of freedom in equal number. We can see that the bosonic fields
$D,D^{*},f^{I},f^{*I},h,h^{*},g^{IJ}$ and $g^{*IJ}$ works as
auxiliary fields. The bosonic fields $s,s^{*},R_{\mu \nu
}^{I},\rho
^{I},\rho ^{*I},r^{IJ},r^{*IJ}$ and the fermionic fields $\xi ,\bar{\xi}%
,\tau ^{I},\bar{\tau}^{I},F^{I},\bar{F}^{I},\zeta
^{IJ},\bar{\zeta}^{IJ}$ work as background fields breaking the
Lorentz invariance.

\section{Concluding Remarks and Comments}

In the important context of studying the gauge invariant Lorentz-violating term
formulated as a Chern-Simons action term , we propose here its $N=2$ and $N=4$
supersymmetric versions. This programme could be done in a simple way with the help of a
dimensional reduction method; here, we have chosen the method \`{a} la Scherk,
but it would also be interesting to contemplate other possibilities, such  as the procedures  \`{a} la Legendre or
\`{a} la Kaluza-Klein. With our reduction scheme,
we could treat the extended supersymmetric version in terms of simple $%
N=1 $ superspace to supersymmetrize the Chern-Simons like term, as
proposed by Jackiw, written in terms of a constant background vector
here parametrized as the
gradient of the scalar function $\alpha +\beta _{\mu }x^{\mu },$ where $%
\alpha $ and $\beta ^{\mu }$ are constants.

Another interesting point we should consider is the possibility, once we have now
the full set of SUSY partners of the Lorentz-breaking vector, to express the central
charges of the extended models whenever topologically non-trivial configurations are
taken into account. This would allow us to impose bounds on the central charges in
terms of the phenomenological constraints imposed on the vector responsible for the
Lorentz covariance breakdown.

\section{Appendix}

The $\sigma ^{\mu }$ and $\bar{\sigma}^{\mu }$ matrices are
defined as
\[
\sigma ^{\mu }=(1,\sigma ^{i}),\,\,\,\,\,\,\,\,\,\bar{\sigma}^{\mu
}=(1,-\sigma ^{i}),
\]
where $\sigma ^{i}$ are the Pauli matrices.

The $SO(3,1)$ generators are represented by the matrices
\[
\sigma ^{\mu \nu }=\frac{i}{4}(\sigma ^{\mu }\bar{\sigma}^{\nu
}-\sigma ^{\nu }\bar{\sigma}^{\mu }).
\]
A useful relation envolving $\sigma $ matrices is
\[
\sigma _{a\dot{a}}^{\mu }\bar{\sigma}^{\nu \dot{a}b}\sigma _{b\dot{b}%
}^{\kappa }\bar{\sigma}^{\lambda \dot{b}a}=2(\eta ^{\mu \nu }\eta
^{\kappa \lambda }-\eta ^{\mu \kappa }\eta ^{\nu \lambda }+\eta
^{\mu \lambda }\eta ^{\nu \kappa }-i\varepsilon ^{\mu \nu \kappa
\lambda }),
\]
where $\varepsilon ^{0123}=-\varepsilon _{0123}=1.$

The Grassmmanian coordinates $\theta $ and $\bar{\theta}$ have
their indices lowered and raised as
\[
\theta ^{a}=\varepsilon ^{ab}\theta _{b},\,\,\,\,\,\,\,\,\,\theta
_{a}=\varepsilon _{ab}\theta ^{b},\,\,\,\,\,\,\,\,\,\bar{\theta}^{\dot{a}%
}=\varepsilon ^{\dot{b}}\bar{\theta}_{\dot{b}},\,\,\,\,\,\,\,\,\,\bar{\theta}%
_{\dot{a}}=\varepsilon _{\dot{a}\dot{b}}\bar{\theta}^{\dot{b}},
\]
where $\varepsilon ^{12}=\varepsilon _{21}=\varepsilon ^{\dot{1}\dot{2}%
}=\varepsilon _{\dot{2}\dot{1}}=1$ and $\varepsilon
^{ab}=-\varepsilon
^{ba},\,\varepsilon _{ab}=-\varepsilon _{ba},\,\varepsilon ^{\dot{a}\dot{b}%
}=-\varepsilon ^{\dot{b}\dot{a}},\,\varepsilon _{\dot{a}\dot{b}%
}=-\varepsilon _{\dot{b}\dot{a}}.$

The covariant derivative in the superspace is
\[
D_{a}=\partial _{a}+i\sigma _{a\dot{a}}^{\mu
}\bar{\theta}^{\dot{a}}\partial
_{\mu },\,\,\,\,\,\,\,\,\,\,\bar{D}_{\dot{a}}=-\bar{\partial}_{\dot{a}%
}-i\theta ^{a}\sigma _{a\dot{a}}^{\mu }\partial _{\mu }.
\]



\begin{thebibliography}{99}
\bibitem{1}  S. Carroll, G. Field and R. Jackiw, {\it Phys. Rev.}{\bf \ D 41}%
, 1231 (1990).

\bibitem{1b}  For an overview see, e.g., {\it CPT and Lorentz Symmetry II},
edited by V.A. Kosteleck\'{y} (World Scientific, Singapore,2002).

\bibitem{1c}  V.A. Kosteleck\'{y}, {\it Phys. Rev.} {\bf D 69}, 105009
(2004).

\bibitem{1d}  V.A. Kosteleck\'{y} and R. Lehnert, {\it Phys. Rev.} {\bf D 63}%
, 065008 (2001).

\bibitem{1e}  R. Lehnert and R. Potting, hep-ph/0408285 (2004)

\bibitem{3a}  M. S. Berger and V. A. Kostelecky, {\it Phys. Rev.} {\bf D 65}%
, 091701 (2002); M. S. Berger, {\it Phys. Rev.} {\bf D 68}. 115005
(2003).

\bibitem{3}  H.J. Belich, J.L. Boldo, L.D. Colatto, J.A. Hela\"{y}el-Neto
and A.L.M.A. Nogueira, {\it Phys. Rev.} {\bf D 68} 065030 (2003).

\bibitem{3c}  A.P. Ba\^{e}ta Scarpelli, H.J. Belich, J.L. Boldo, L.D.
Colatto, J.A. Hela\"{y}el-Neto and A.L.M.A. Nogueira, {\it Nucl. Phys.} {\bf %
B}- {\it Supp}.127, 105 (2004).

\bibitem{3b}  S. G. Nibbelink and M. Pospelov, hep-ph/0404271 (2004).

\bibitem{2}  S. J. Gates Jr., M. T. Grisaru, M. Rocek and W. Siegel,
Superspace or One Thousand and one Lessons in Supersymmetry
(Benjamin, Massachusetts, 1983).

\bibitem{2b}  L. Brink, J. Schwarz and J. Scherk, {\it Nucl. Phys}. {\bf B}
121, 77 (1977)

\bibitem{2c}  P. Fayet, {\it Nucl. Phys}. {\bf B}
113, 135 (1976)

\bibitem{2d}  F. Gliozzi, J. Scherk and D. Olive, {\it Nucl. Phys}. {\bf B}
122, 253 (1977)

\bibitem{5}  V. E. R. Lemes, A.L.M.A. Nogueira and J. A. Hela\"{y}el-Neto,
{\it International Journal of Modern Physics} {\bf A}, Vol. 13,
No. 18 (1998) 3145.
\end{thebibliography}
\end{document}